\RequirePackage{fix-cm}
\documentclass{svjour3}                     
\smartqed  
\usepackage{graphicx}
\usepackage{color}
\usepackage{amsmath,amssymb}
\usepackage{bm}
\usepackage{epsfig}
\usepackage{citesort}
\usepackage[neverdecrease]{paralist}
\usepackage{url}
\usepackage{epstopdf}
\usepackage{setspace}
\usepackage {color}

\begin{document}

\title{Ionic Coulomb blockade and anomalous mole fraction effect in NaChBac bacterial ion channels}

\titlerunning{Ionic Coulomb blockade and AMFE in NaChBac bacterial ion channels}        

\author{   I.Kh. Kaufman     \and
           O.A. Fedorenko    \and  \\
           D.G. Luchinsky    \and
           W.A.T. Gibby       \and   \\
           S.K. Roberts      \and
           P.V.E. McClintock \and   \\
           R.S. Eisenberg
}


\institute{I.Kh. Kaufman, W.A.T. Gibby, P.V.E. McClintock \at Department of Physics, Lancaster University, Lancaster LA1 4YB, UK
           \\ \email{p.v.e.mcclintock@lancaster.ac.uk.}       
           \and O.A. Fedorenko, S.K. Roberts \at Division of Biology and Life Sciences, Lancaster University, Lancaster UK
           \and D.G. Luchinsky  \at Department of Physics, Lancaster University, Lancaster LA1 4YB, UK and SGT, Inc., Greenbelt, MD, 20770, USA
           \and R.S. Eisenberg \at Department of Molecular Biophysics, Rush University, Chicago, IL, US
}


\maketitle

\begin{abstract}
We report an experimental study of the influences of the fixed charge and bulk ionic concentrations on the conduction of biological ion channels, and we consider the results within the framework of the ionic Coulomb blockade model of permeation and selectivity. Voltage clamp recordings were used to investigate the Na$^+$/Ca$^{2+}$ anomalous mole fraction effect (AMFE) exhibited by the bacterial sodium channel NaChBac and its mutants. Site-directed mutagenesis was used to study the effect of either increasing or decreasing the fixed charge in their selectivity filters for comparison with the predictions of the Coulomb blockade model. The model was found to describe well some aspects of the experimental (divalent blockade and AMFE) and simulated (discrete multi-ion conduction and occupancy band) phenomena, including a concentration-dependent shift of the Coulomb staircase. These results substantially extend the understanding of ion channel selectivity and may also be applicable to biomimetic nanopores with charged walls.


\end{abstract}


\section {Introduction}
Biological ion channels are natural nanopores providing for the fast and highly selective permeation of physiologically important ions (e.g.\ Na$^+$, K$^+$ and Ca$^{2+}$) through cellular membranes \cite{Hille:01,Zheng:15,Eisenberg:13a}. Despite their fundamental importance, and notwithstanding enormous efforts by numerous scientists, the physical origins of their selectivity
still remain unclear. It is known, however, that the conduction and selectivity properties of cation channels are defined by the ions' movements and interactions inside a short, narrow selectivity filter (SF) lined by negatively charged amino acid residues that provide a net fixed charge $Q_f$ \cite{Hille:01,Zheng:15}.

NaChBac bacterial sodium channels \cite {Yue:02,Tang:16,Naylor:16} are frequently thought of, and used as, simplified experimental/simulation models of mammalian calcium and sodium channels. X-ray investigation and molecular dynamics simulations have shown that these tetrameric channels possess strong binding sites with 4-glutamate \{EEEE\} loci at the SF \cite{Payandeh:12}. Bacterial channels have been used in site-directed mutagenesis (SDM) /patch clamp studies of conductivity and selectivity \cite {Corry:12,Naylor:16}.

Conduction and selectivity in calcium/sodium ion channels have  recently been described \cite {Kaufman:13a,Kaufman:13b,Kaufman:15} in terms of ionic Coulomb blockade (ICB) \cite {Krems:13,Feng:16}, a fundamental electrostatic phenomenon based on charge discreteness, an electrostatic exclusion principle, and single-file stochastic ion motion through the channel. Earlier, Von Kitzing had revealed the staircase-like shape of the occupancy {\it vs} site affinity for the charged ion channel \cite {Kitzing:92} (following discussions and suggestions  in \cite {Eisenberg:96}), and comparable low-barrier ion-exchange transitions had been discovered analytically \cite {Zhang:06}. A Fermi distribution of spherical ions was used as the foundation of a Poisson Fermi theory of correlated  ions in channels \cite {Liu:13a,Liu:14}.

ICB has recently been observed in sub-nm nanopores \cite {Feng:16}. It appears to be closely similar to its electronic counterpart in quantum dots \cite {Beenakker:91}. As we have demonstrated earlier \cite {Kaufman:15}, strong ICB appears for Ca$^{2+}$ ions in model biological channels and manifests itself as an oscillation of the conductance as a function of $Q_f$, divalent blockade, and AMFE.

Here we present an enhanced model of divalent blockade and AMFE able to encompass concentration-related shifts in the ICB conduction bands and the shape of the divalent blockade decay. We compare model predictions with the literature, with our own earlier simulated data \cite {Kaufman:13b,Kaufman:15}, and with new experimental results from a patch clamp study of conductivity and selectivity in the NaChBac bacterial channel and its mutants that has enabled $Q_f$ to be changed. ICB-based $Q_f$ {\it vs.} log$[Ca]$ phase diagrams are introduced to explain visually the differences between the AMFE behaviours observed for different mutants, where $[Ca]$ is the Ca$^{2+}$ ion concentration.

In what follows $\varepsilon_0$ is the permittivity of free space, $e$ is the proton charge, $z$ is the ionic valence, $T$ the temperature and  $k_B$ is Boltzmann's constant.  

\section {Models and methods}
\begin{figure}[h!]
\includegraphics[width=0.8\textwidth]{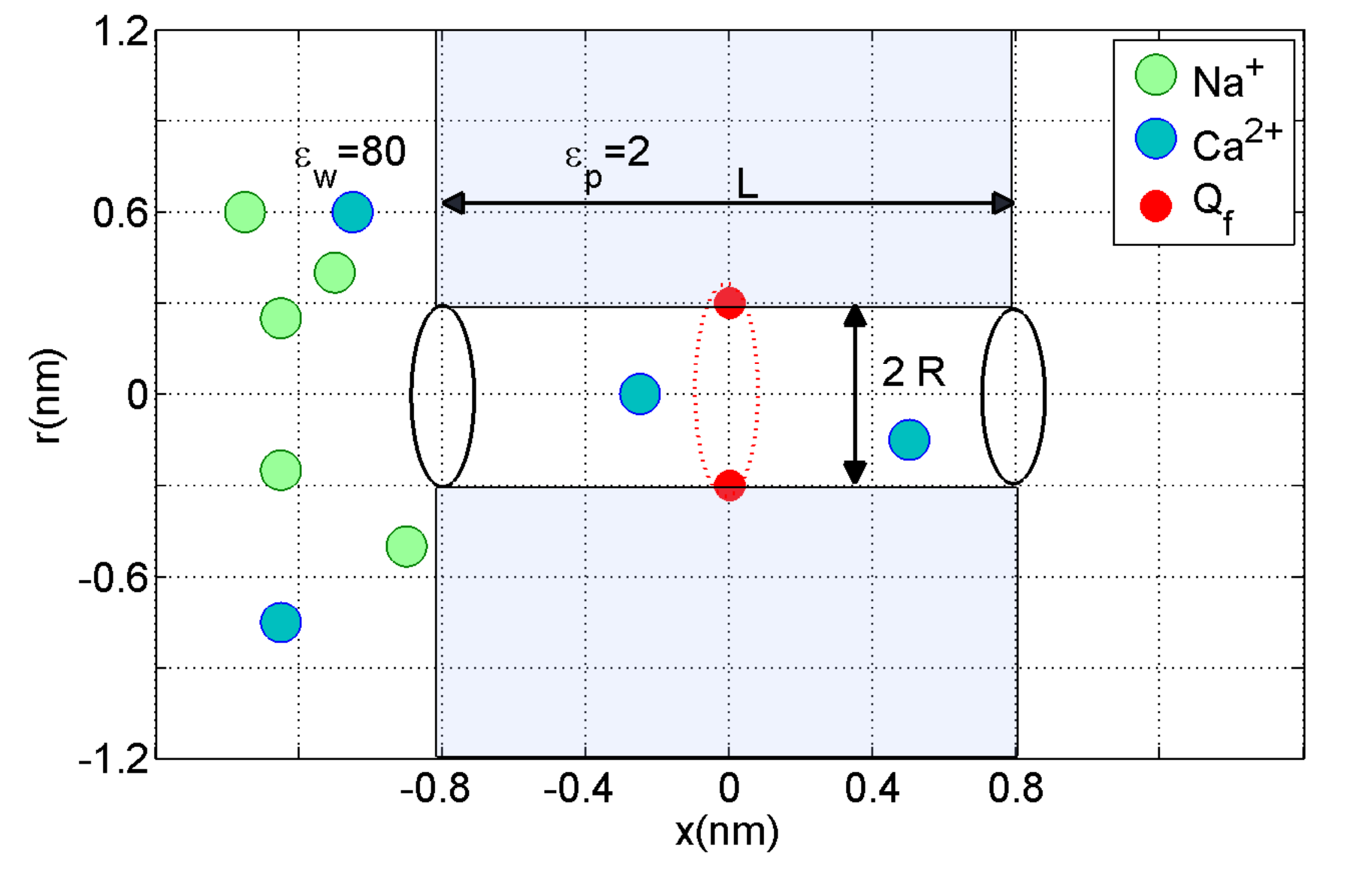}
\caption{Generic electrostatic model of Calcium/Sodium ion channel \cite {Kaufman:13b}. The model describes the selectivity filter of ion channel as an axisymmetric, water-filled pore of radius $R=0.3$nm ~and length $L=1.6$\,nm through the cellular membrane. A centrally-placed, uniform, rigid ring of negative charge $Q_f$ is embedded in the wall to represent the charged residues of real Ca$^{2+}$/Na$^+$ channels. We take both the water and the protein to be homogeneous continua describable by relative permittivities $\varepsilon_w=80$ and $\varepsilon_p=2$, respectively, together with an implicit model of ion hydration whose validity is discussed elsewhere. The moving monovalent Na$^{+}$ and divalent Ca$^{2+}$ ions are assumed to obey self-consistently both Poisson's electrostatic equation and the Langevin equation of motion.
}
\label{fig:channel}
\end{figure}

Figure \ref{fig:channel} summarises the generic, self-consistent, electrostatic model of the selectivity filter of a calcium/sodium channel introduced earlier \cite {Kaufman:13b}. It consists of a negatively-charged, axisymmetric,  water-filled, cylindrical pore through the protein hub in the cellular membrane; and, we suppose it to be of radius $R=0.3$\,nm and length $L=1.6$\,nm \cite {Corry:01}, to match the dimensions of the selectivity filters of Na$^+$/Ca$^{2+}$ channels.

There is a centrally-placed, uniformly-charged, rigid ring of negative charge $0 \leq |Q_f/e| \leq 8$ embedded in the wall at $R_Q=R$ to represent the charged protein residues of real Ca$^{2+}$/Na$^+$ channels. The left-hand bath, modeling the extracellular space, contains non-zero concentrations of Ca$^{2+}$ and/or Na$^+$ ions. For the Brownian dynamics simulations, we used a computational domain length of $L_d=10$\,nm and radius $R_d=10$\,nm, a grid size of $h=0.05$\,nm, and a potential difference in the range $0-25$\,mV (corresponding to the depolarized membrane state) was applied between the left and right domain boundaries. We take both the water and the protein to be homogeneous continua describable by relative permittivities $\varepsilon_w=80$ and $\varepsilon_p=2$, respectively, together with an implicit model of ion hydration whose validity is discussed elsewhere \cite{Kaufman:13b}.

Of course, our reduced model represents a significant simplification of the actual electrostatics and dynamics of ions and water molecules within the narrow selectivity filter due to, for example: the application of continuum electrostatics; the use of the implicit solvent model;  and the assumption of 1D (i.e.\ single-file) movement of ions inside the selectivity filter. The validity and range of applicability of this kind of model have been discussed in detail elsewhere \cite{Kaufman:13b,Kaufman:15,Roux:04}.

This simplified self-consistent model was used as the basis for development of the ICB model of permeation and selectivity \cite{Kaufman:15,Kaufman:13b}, which led to some predictions that we now test experimentally in two complementary ways: through site-directed mutagenesis and patch-clamp studies of the bacterial sodium NaChBac channel; and numerically through Brownian dynamics simulations.

The main aim was to test the ICB model's predictions of the dependence of the conductivity type, and of the divalent blockade/AMFE properties, on the fixed charge $Q_f$ at the SF. Site-directed mutagenesis and patch clamp measurements were used to investigate changes in the ion transport properties of NaChBac mutants caused by alterations in the amino acid residues forming the SF, i.e.\ alterations in $Q_f$. Increasing the value of $Q_f$ was expected to lead to stronger divalent blockade following the Langmuir isotherm and to a resonant variation of the divalent current with $Q_f$ \cite{Kaufman:15}.

The SF of NaChBac is formed by 4 trans-membrane segments each containing the six-amino-acid sequence LESWAS (leucine/glutamic-acid/serine/tryptophan/\\alanine/serine, respectively, corresponding to residues 190 –- 195).
This structure provides the highly-conserved \{EEEE\} locus with a nominal $Q_f=-4e$ \cite {Yue:02}. As summarised in Table 1, site-directed mutagenesis was used to generate two mutant channels in which the SF either has ``deleted charge'' $Q_f=0$ (L\textbf{A}SWAS, in which the negatively charged glutamate E191 is replaced with electrically neutral alanine)  or has ``added charge'' $Q_f=-8e$ (LE\textbf{D}WAS, in which the electrically neutral serine S192 is replaced by negatively charged aspartate D \cite {Yue:02}.

Details of the methods used for preparation of the mutants, and for the electro-physiology measurements, are presented in   Appendix 1.

\begin{table}[t!]
\caption{Main properties of the wild type (LESWAS) and mutant (LASWAS and LEDWAS) NaChBac bacterial channels generated and used for the present patch-clamp study. Here $Q_f$ stands for the nominal fixed charge at the selectivity filter and IC$_{50}$ is the the [$Ca$] threshold value providing 50\% blockade of the Na$^+$ current. Qualitative properties (selectivity and AMFE) are marked as ``+'' where present and ``$-$'' where absent. }
\begin{tabular}{|p{2.1cm}|p{2.2cm}|p{1.3cm}|p{2.0cm}|p{2.4cm}|}\hline
Channel &  SF amino acid sequence & Nominal $Q_f/e$ & Ca/Na selectivity & Ca/Na AMFE \\
\hline
NaChBac wild-type& L\textbf{ES}WAS & $-4$ & + (Na$>$Ca) & $-$  \\
   & & & & \\
Zero-charge mutant& L\textbf{\underline{A}}SWAS & ~~0 & $-$ & $-$  \\
   & & & & \\
Added-charge mutant & LE\textbf{\underline{D}}WAS & $-8$ & + (Ca$>$Na) & + (IC$_{50}$=5$\mu$M) \\
\hline
\end{tabular}
\label{tab:mutants}

\end{table}

\section {Results and Discussion}
\begin{figure}[t!]
\includegraphics[width=1.0\linewidth]{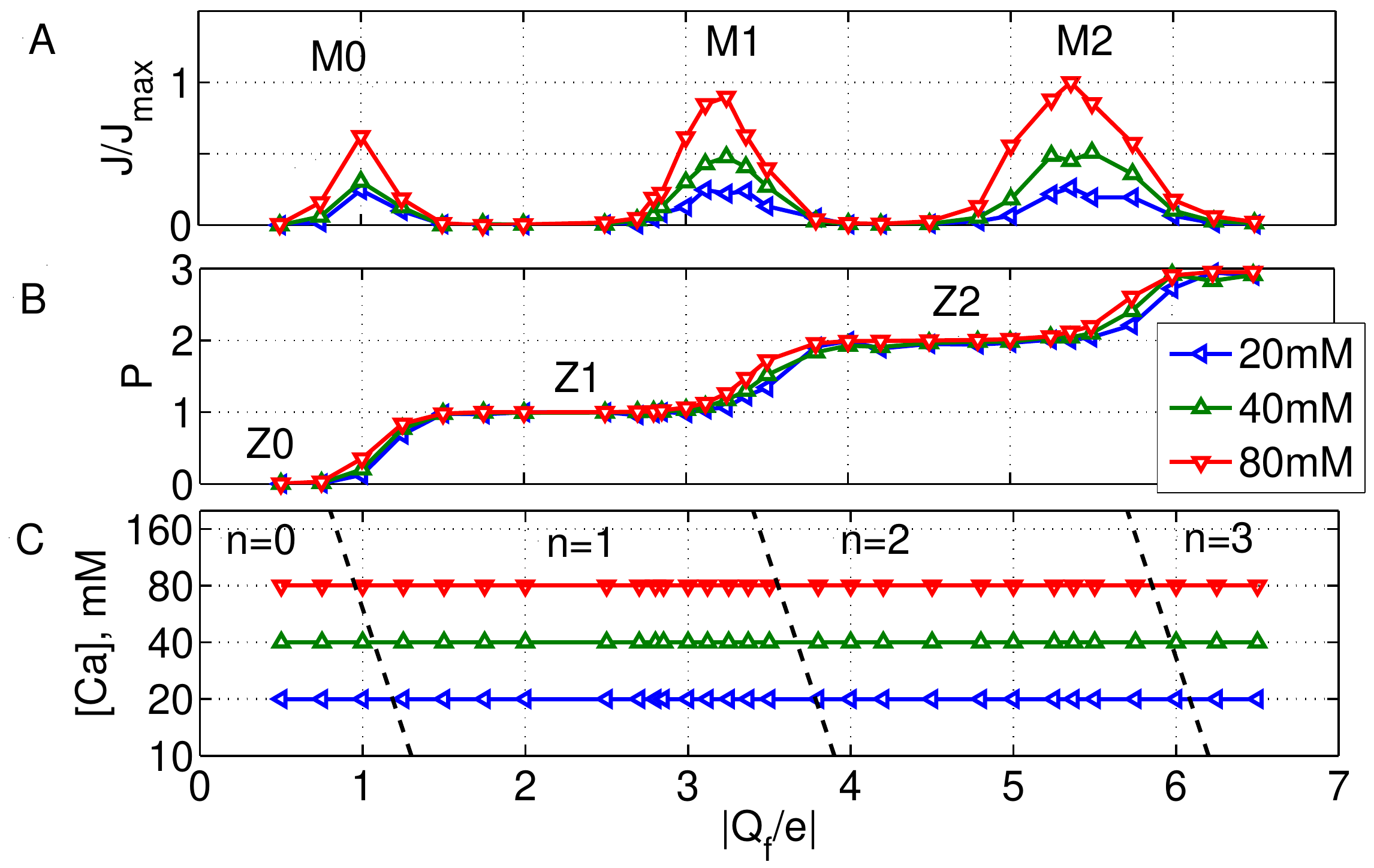}
\caption{Multi-ion conduction/occupancy bands in the model calcium channel, showing occupancy shifts with ionic concentration. A. Multi-ion calcium conduction bands $M_n$ as established by Brownian dynamics simulations. B. The corresponding Coulomb staircase of occupancy $P_c$ for different values of the extracellular calcium concentration $[Ca]$, as marked, consists of steps in occupancy as $[Ca]$ changes. The neutralized states $Z_n$ providing blockade are interleaved with resonant states $M_n$. (Plots A, B are taken from \cite {Kaufman:13b}) C. Coulomb blockade-based phase diagram. The positions of the $\{n\}\rightarrow\{n+1\}$ transitions  (from equation (\ref {equ:ICB_shift})) are shown as black dashed lines. The horizontal coloured lines are guides to the eye, indicating the three concentrations used in the simulations. The diagram is consistent with the logarithmic $[Ca]$-related shift of steps in the Coulomb staircase shown in B.
} \label{fig:ICB_bands}
\end{figure}

Coulomb blockade (whether ionic or electronic) arises in low-capacitance, discrete-state systems for which the ground state \{$n_G$\} with $n_G$ ions in the channel is separated from neighbouring $\{n_G\pm 1\}$ states by a deep Coulomb gap $U_{s} \gg k_B T$, so that we can define the strength of the ICB as $S_{ICB}=U_s/(k_B T)$. The ICB phenomenon manifests itself as multi-ion oscillations (alternating conduction bands and stop bands) in the Ca$^{2+}$ conductance and channel occupancy \cite {Kaufman:13a,Kaufman:15}.

Figure \ref {fig:ICB_bands} summarises the results of our earlier \cite{Kaufman:13a}  Brownian dynamics simulation of Ca$^{2+}$ conduction, which was found to occur in multi-ion bands: A shows strong oscillations of conductance; and B the corresponding occupancy $P$, which was found to take the form of a Coulomb staircase where the steps in $P$ occur in between resonant conduction points $M_n$ and current blockade points $Z_n$, as predicted by the ICB-based linear response model. Closer inspection of figure \ref {fig:ICB_bands}B shows that the Coulomb staircase exhibits small concentration-related shifts \cite {Kaufman:15}. We now present a simple model to account for this shift, leading to the phase diagram shown in Figure \ref {fig:ICB_bands}C.

We define the positions of the  resonant conduction $M_n$ points (where barrier-less conduction can occur because $G_n = G_{n+1}$, where $G_n$ is the Gibbs free energy when there are $n$ ions in the SF) taking account of concentrations $P_b$ and $P_c$.  We assume that the blockade is strong and we approximate $U_n$ by the dielectric self-energy of the excess charge $Q_n$ \cite {Kaufman:15}:
\begin{eqnarray}
U_{n} &=& \frac{Q_n^2}{2 C_0} ;  \quad   Q_n = z e n +Q_f; \quad
C_0  = \frac {4 \pi \varepsilon_0 \varepsilon_w R^2}{ L}
\label {equ:excess}
\end{eqnarray}
Here, $C_0$ stands for the geometry-dependent self-capacitance of the channel .

In equilibrium, the chemical potentials in the bulk $\mu_b$ and in the channels $\mu_c$ are equal\cite {Nonner:01,Krauss:11}:
\begin {eqnarray}
\mu_b&=&\mu_{b,0} + k_B T \ln P_b \quad \mbox{(Chemical potential of ions in the bath)} \\
\mu_c&=&\mu_{c,0} + \Delta \mu_{c,ex} \quad\quad\: \mbox{(Chemical potential of ions in the SF)}  \\
\mu& =& \mu_b = \mu_c \hspace*{1.65cm} \mbox {(Equibrium condition)}
\end {eqnarray}
where the standard potentials $\mu_{b,0}$ and $\mu_{c,0}$ are assumed to be zero (although other choices are possible, \cite {Kaufman:15}), $P_b$ and $P_c=\langle n \rangle $ stand for equivalent bulk, and the SF occupancy is related to the SF volume $V_{SF}=\pi R^2 L$, i.e. $P_b=n_b V_{SF}$, where $n_b$ is bulk number density of the species of interest.

The excess chemical potential in the SF, $\Delta \mu_{c,ex}$, is defined  here as the excess Gibbs free energy $\Delta G_n=\Delta U_n - T \Delta S_n$ in the SF due to the single-ion $\{n\} \rightarrow \{n+1\}$ transition, from (\ref {equ:excess}). The SF entropy-related term $T \Delta S_n$ is model-dependent, varying between the extremes for correlated motion and for an ideal gas \cite {Nonner:01,Luchinsky:16}. We use the ``single-vacancy'' model of the motion \cite {Nelson:11,Kaufman:15} for which the following result can be derived \cite{Luchinsky:16}:
\begin {equation}
\Delta G_n=\Delta U_n +k_B T\ln(n+1)
\end {equation}
Hence, the equilibrium ($\mu_b=\mu_c$) occupancy around the transition point $M_n$  represents a thermally-rounded staircase (see Figure \ref{fig:ICB_bands}B)  described by a Fermi-Dirac distribution \cite {Kaufman:15,Zhang:06}:
\begin{equation}
P^*_c=P_c-n=
\left [ 1+  \exp \left(\frac{ (\Delta G_n-\mu_b)}{k_B T}\right)\right]^{-1}
=\left[ 1+ \frac{1}{P_b}\exp \left(\frac{\Delta G_n}{k_B T} \right) \right]^{-1}
\label {equ:fd_qf}
\end{equation}
It corresponds to the Coulomb staircase (Figure \ref {fig:ICB_bands}B), well-known in Coulomb blockade theory \cite {Beenakker:91} which  appears when varying either $Q_f$ or $\log (P_b)$.

The resonant value $ M_n$ of $ Q_f$ for the $\{n\} \rightarrow \{n+1\}$ transition is defined as:
\begin{eqnarray}
M_n & = & - z  e (n + 1/2) - \delta  M_n; \hspace*{2.2cm} \mbox{(Nominal transition point)}    \\
\label {equ:ICB_shift0}
\delta M_n &=& z e  \frac {C_0 k_B T}{z^2 e^2} \left[\ln (P_b) -\ln(n+1) \right] \hspace*{1.0cm} \mbox{(Concentration-related shift)}
\label {equ:ICB_shift}
\end{eqnarray}
Next we introduce the notion of ``phase diagrams'' and use them to describe the concentration-related shifts seen in our earlier Brownian dynamics simulations (Figure \ref{fig:ICB_bands}B) and the divalent blockade/AMFE in mutation experiments on the bacterial NaChBac channel that we report below.

\begin{figure}[t!]
\includegraphics[width=1.0\textwidth]{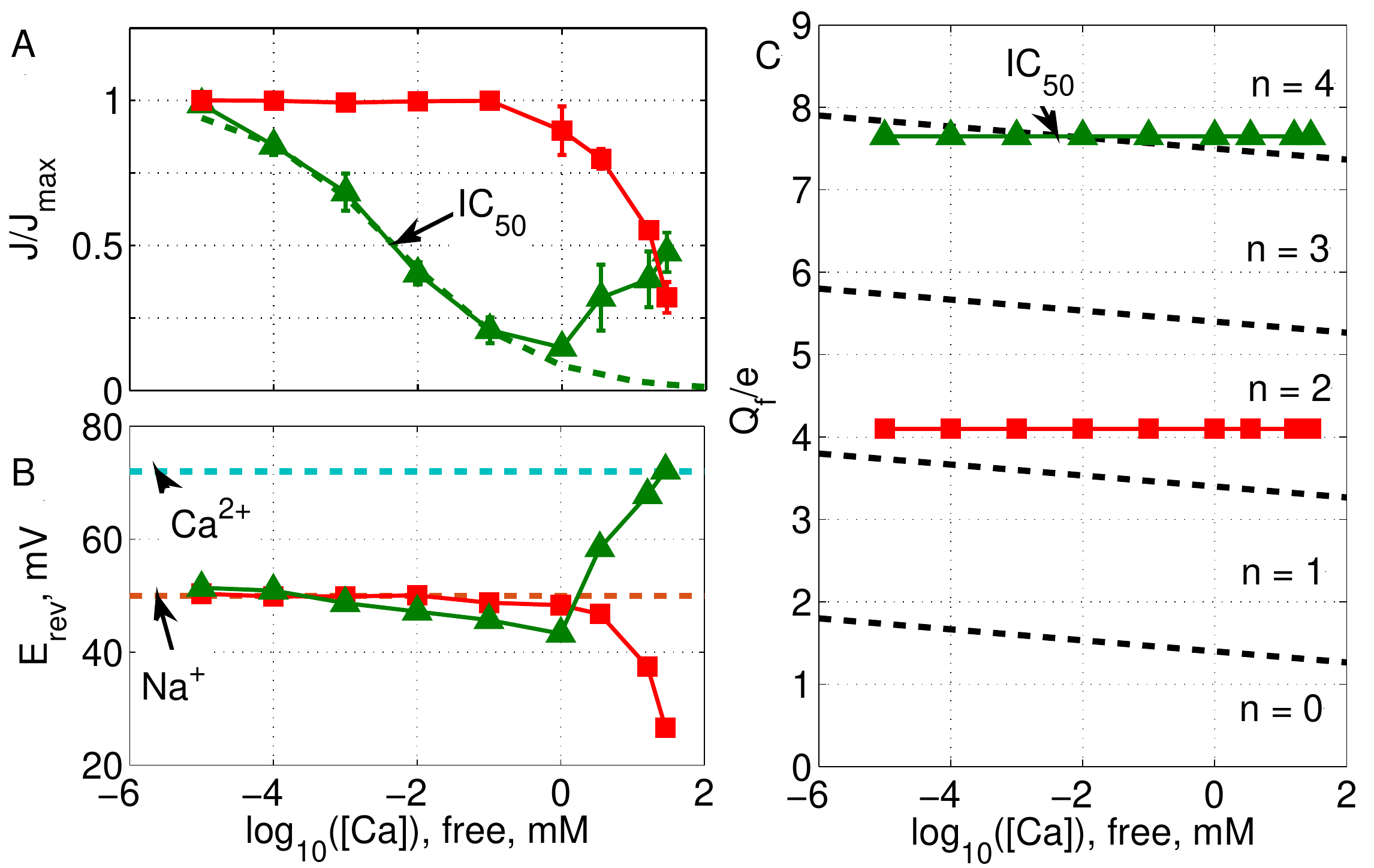}
\caption{Divalent blockade and anomalous mole fraction effect (AMFE) in wild-type NaChBac (LESWAS, shown as red squares) and LEDWAS (green triangles) mutant channels. The bath solution containing Na$^{+}$ and Ca$^{2+}$ cations was adjusted by replacement of Na$^{+}$ with equimolar Ca$^{2+}$; the free Ca$^{2+}$ concentrations $[Ca]$ are shown on the abscissa; and error bars represent the standard error in the mean (SEM).
A: Averaged normalized peak currents of LESWAS  and LEDWAS channels. The Langmuir isotherm (\ref {equ:Langmuir}) fitted to the LEDWAS data for $[Ca] <  1$\,mM is shown by the green dashed line.
B: Reversal potentials ($E_{rev}$) obtained from the same recordings as in A indicate that LEDWAS stopped conducting Na$^{+}$ if [Ca$^{2+}$]$\ge$ 1mM. Dashed lines indicate reference values when extracellular [Ca$^{2+}$] (green) and [Na$^{+}$] (red) are fixed to 100mM.
C:  Cartoon phase diagram $Q_f$ vs log([Ca]), where the switching lines predicted by Equation (\ref{equ:ICB_shift}) are dashed-black.
}
\label{fig:AMFE}       
\end{figure}

The phase diagrams (Figure \ref {fig:ICB_bands}C, Figure \ref {fig:AMFE}C) represent the evolution of the channel state on a 2-D plot with occupation log($P_b$) (or equivalently $\log([Ca])$ concentration) on the ordinate axis and $Q_f/e$ on the abscissa or {\it vice versa}. The phase transition lines (black, dashed) separate  the states of the SF having different integer occupancy number $\{n\}$. Different sections through the diagram reflect different experiments/simulations in the sense that we can choose to vary either the concentration (divalent blockade/AMFE experiments) or $Q_f$ (patch clamp experiments on mutants).

Let start from the concentration-related shift of the Coulomb staircase. Figure \ref {fig:ICB_bands}C shows the switching lines and AMFE trajectory (projection of system evolution) in the Ca$^{2+}$ ionic occupancy phase diagram \cite {Zhang:06} for the calcium/sodium channel while Figure \ref {fig:ICB_bands}B shows the small concentration-related shifts of the Coulomb staircase for occupancy $P$ found in the Brownian dynamics simulations \cite {Kaufman:13a}. Equations (\ref {equ:ICB_shift}) and the phase diagram provide a simple and transparent explanation of the simulation results. The origin of the shift lies in the logarithmic concentration dependence of $\delta M_n$ in (\ref {equ:ICB_shift}). Similar shifts were seen in earlier simulations \cite {Kitzing:92}. Note that the BD simulations seem not to show any significant shift for the $M_0$ points with increasing log($[Ca]$), an unexpected result that requires further investigation.

\begin{figure}[t!]
\includegraphics[width=1.0\textwidth]{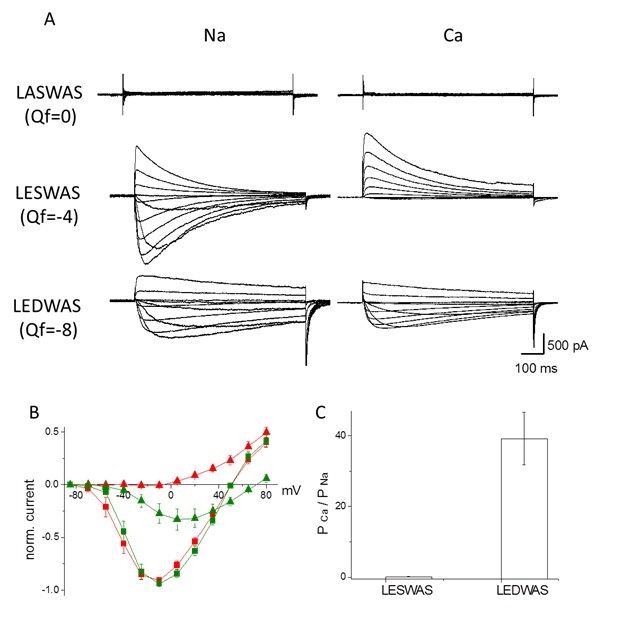}
\caption{Permeability to Na$^+$ and Ca$^{2+}$ of the wild type NaChBac (LESWAS) ion channel and its mutants (LASWAS and LEDWAS). A: Representative whole cell current {\it vs} time records obtained for channels in bath solution containing 140 mM Na$^+$ (left) or 100 mM Ca$^{2+}$ (right). B:  Current-voltage I-V relationships  ($\pm$ SEM are shown as bars, $n=6-12$)) for LESWAS in Na$^+$ solution (red squares) and Ca$^{2+}$ solution (red triangles) and LEDWAS in Na$^+$ solution (green squares) and Ca$^{2+}$ solution (green triangles) normalized to the maximal peak Na$^+$ current from the same cell. C: Permeability Na$^+$/Ca$^{2+}$ ratios determined using reversal potentials, as described in \cite {Sun:97}, indicate that LEDWAS is a Ca$^{2+}$ selective channel.}
\label{fig:traces}       
\end{figure}
Our electrophysiological measurements on NaChBac wild type LESWAS channels and their mutants show what happens in reality. The net currents through a macroscopic number of identical biological channels can be resolved and displayed on a biologically relevant time scale. Figure \ref {fig:traces}A shows the original current traces using bath solutions containing either Na$^+$ or Ca$^{2+}$ as the charge carrying cation  (see Appendix 1 for details of methods used). Zero-charge  mutants ($Q_f=0$) did not show any measurable current in either of the solutions, corresponding well with the Coulomb-blockaded state expected/measured for an uncharged channel or nanopore \cite {Kaufman:15,Feng:16}.

Wild type (LESWAS) channels ($Q_f=-4e$) exhibited high Na$^+$ conductance in agreement with earlier observations \cite {Yue:02,Tang:16} and with the ICB model which predicts relatively $Q_f$-independent Na$^{+}$ conduction due to the small valence $z=1$ of Na$^+$ ions.

LEDWAS channels with nominal $Q_f=-8e$ were found to conduct both Na$^+$ and Ca$^{2+}$ (figure \ref {fig:traces}A).
These results for LESWAS and LEDWAS are consistent with previous reports \cite{Yue:02,Decaen:14,Naylor:16}. They are also in agreement with the Coulomb blockade model, which predicted conduction bands for divalent cations in these mutants.

To study the selectivity between Na$^+$ and Ca$^{2+}$ in more detail and to investigate divalent blockade and AMFE, we performed experiments using bath solutions containing mixtures of Na$^+$ and Ca$^{2+}$, at different concentrations. Ca$^{2+}$ was added
to a bath solution (containing 140 mM Na$^+$) to achieve free Ca$^{2+}$ concentrations from 10\,nM up to 1\,mM, which were calculated by Webmaxc (\url{http://web.stanford.edu/~cpatton/webmaxcS.htm}) and achieved by adding HEDTA (for concentrations from 1 mM to 10 $\mu$M) or EGTA (for concentrations $\leq 1$ $\mu$M).

Figure \ref {fig:AMFE}A also shows that the current through the ``added charge'' mutant channel, LEDWAS, was highly sensitive to the presence of Ca$^{2+}$, and fell rapidly with increasing $[Ca]$, i.e.\ it exhibited strong Ca$^{2+}$ blockade of its Na$^+$ currents. This first part of the AMFE phenomenon is well-known for calcium channels as divalent blockade \cite {Sather:03,Tang:16}.  The blockade shape is frequently fitted empirically with a Langmuir isotherm, similarly to the cases of blockade by dedicated channel blocker drugs \cite {Tang:16}.

A complete description of divalent blockade and AMFE should account for statistical and kinetic features of the multi-species solution inside the SF \cite {Luchinsky:16,Liu:13a}. We use a simplified description based on the assumptions:
\begin {equation}
[Ca]\ll[Na]; \quad \tau_{Ca}\gg \tau_{Na}; \quad P_c([Ca])+P_c([Na]) \le 1
\end {equation}
where $\tau_{Ca}$ and $\tau_{Na}$ stand for the respective ionic binding times. Under these assumptions, the SF can be in two exclusive states: ``open'' ($P_c([Ca])=0,$ $J_{[Na]}([Ca])=J_{[Na]}(0)$); and ``closed'', blocked by Ca$^{2+}$ ions, $(P_c([Ca])=1$, $J_{[Na]}([Ca])=0$), and the states are shared in time. Hence due to the ergodic hypothesis the blockade of the Na$^+$ current reflects Ca$^{2+}$ occupancy:
\begin{equation}
J_{Na}([Ca])=J_{Na}(0) (1-P_c(Ca))
\end{equation}
The ICB model \cite {Kaufman:15} {\it predicts} that  blockade by Ca$^{2+}$ (or any other strong blocker) can be described by the Langmuir isotherm:
\begin{equation}
    X([Ca])=\ln \left(\frac{J([Ca])}{J(0)-J([Ca])}\right)= \ln \left(\frac{1-P^*_c}{P^*_c}\right)=\ln (IC_{50}) - \ln ([Ca])
\label {equ:Langmuir}
\end{equation}
where the monovalent partial current $J([Ca])$ as a function of the bulk concentration $[Ca]$) is described by a Fermi-Dirac function (\ref {equ:fd_qf}) that is equivalent to the Langmuir isotherm (\ref {equ:Langmuir}) and $IC_{50}$ is Ca$^{2+}$ concentration when $J(IC_{50})=0.5 J(0)$.  Note that (\ref {equ:Langmuir}) strictly predicts a logarithmic slope of unity, $s=dX/d\ln([Ca])=1$. A similar equation was derived in \cite{Liu:14}.

Figure \ref {fig:AMFE}A demonstrates an absence of divalent blockade for LESWAS in marked contrast with the strong blockade for LEDWAS mutants, which is well-fitted by the Langmuir isotherm (\ref {equ:Langmuir}) with a threshold value $IC_{50}= 5\mu$M ($s=1\pm 0.05$). The $IC_{50}$ value can in principle be connected to $Q_f$ \cite{Kaufman:15} but it will require better knowledge of the SF dimensions and will be a target of future research.

Figure \ref {fig:AMFE}B shows that $E_{rev}$ for LEDWAS mutant starts from the same 50mM (the value measured for a 100mM Na$^+$ bath) as LESWAS but, from the point where the current starts to increase with growth of $[Ca]$ ($\approx$1mM), it rises rapidly to 72mM, which is the value measured for a 100mM  Ca$^{2+}$ bath. It implies that, similarly to the L-type calcium channel, AMFE in the LEDWAS mutant involves the substitution of the sodium current by the calcium one. Equation (\ref {equ:Langmuir}) is also applicable to the drug-driven blockade of bacterial mutants \cite{Tang:16}.

Taken together, the results described above provide some experimental validation of the ICB model. In particular, they confirm the importance of $Q_f$ as a determinant of NaChBac ionic valence selectivity. Increasing the negative charge in the SF results in permeability for divalent cations and it leads to phenomena such as divalent blockade of the Na$^+$ current and AMFE. Moreover, the close fitting of the current decay by (\ref {equ:Langmuir}) confirms one of the main ICB results, {\it viz.} that the SF occupancy is described by a Fermi-Dirac distribution.
\begin{figure}[h!]
\includegraphics[width=1.0\linewidth]{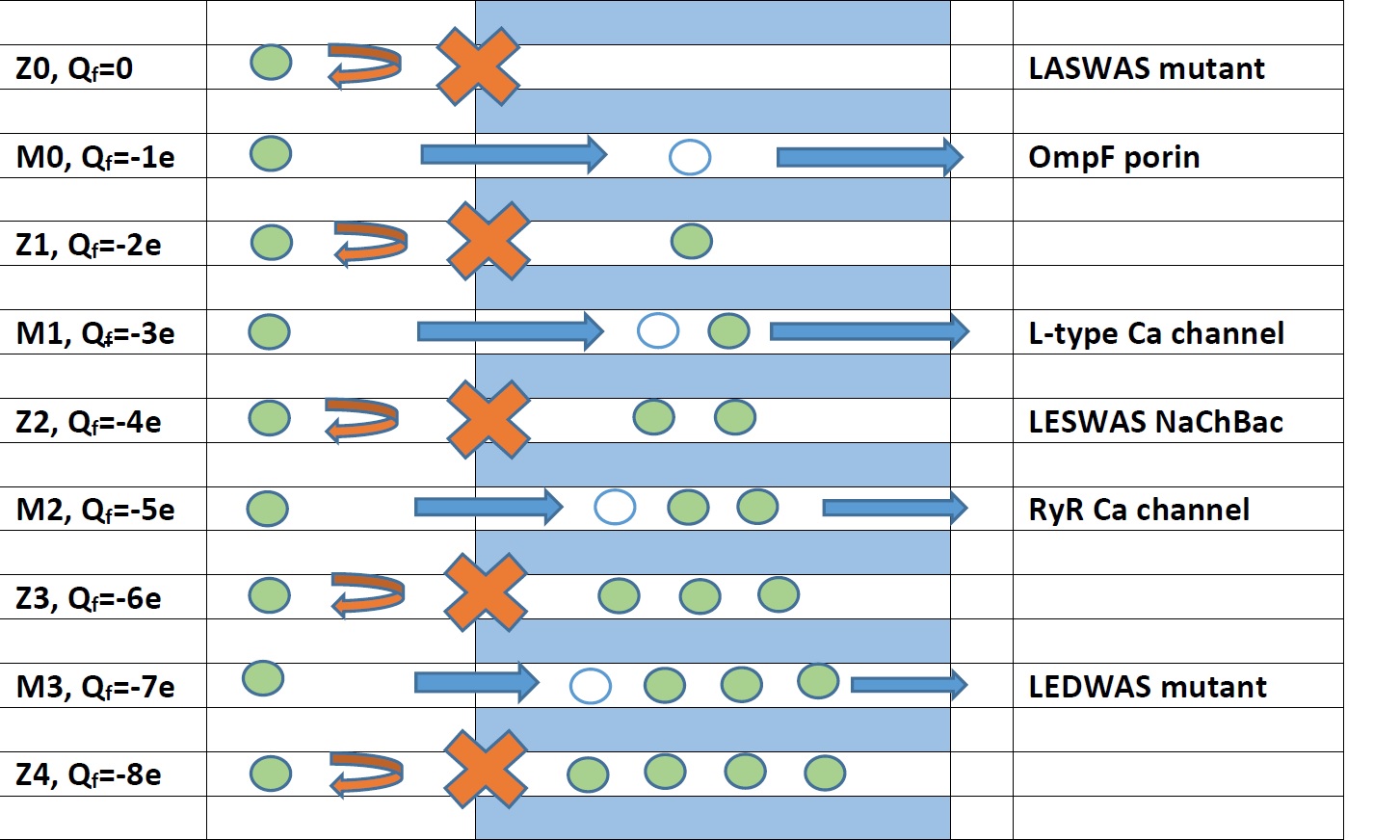}
\caption{Evolution of the Ca$^{2+}$ conduction mechanism with increasing absolute value of effective fixed charge $|Q_f|$, showing the Coulomb blockade oscillations of multi-ion conduction/blockade states.  The neutralized states $Z_n$ providing blockade are interleaved with resonant conduction states $M_n$. The $|Q_f|$ value increases from top to bottom, as shown. Green circles indicate Ca$^{2+}$ ions, unfilled circles show vacancies (virtual empty states during the knock-on process). The right-hand column indicates the preliminary identifications of particular channels/mutants corresponding to particular mechanisms.
}
\label{fig:ICB_scheme}
\end{figure}

Figure \ref {fig:ICB_scheme} illustrates diagrammatically the quasi-periodic $\{n\}$--sequence of multi-ion blockade/conduction modes described by (\ref {equ:ICB_shift}) with growth of $\{n\}$, where $Q_f$ or $P_b$ increase, together with putative identifications of particular modes. The state with $Z_0=0$ represents ionic Coulomb blockade of the ions  at the selectivity filter  by image forces -- as observed experimentally in LASWAS (see above) and also in artificial nanopores \cite{Feng:16}. The first resonant point $M_0$ corresponds to single-ion (i.e.\ $\{n\}=\frac{1}{2}$) barrier-less conduction, and can be related to the OmpF porin \cite {Miedema:04,Kaufman:13b}.  This state is followed by $Z_1$ and $M_1$ states describing double-ion knock-on and identified with L-type calcium channels \cite {Kaufman:13a}. The three-ion resonance $M_3$ can be identified with the RyR calcium channels \cite {Gillespie:08}. On a preliminary basis, NaChBac channels can be identified with the $Z_2$ blockade point, and their LEDWAS mutant with the calcium-selective $M_3$ resonant point. Further molecular dynamics simulations will be needed to resolve the observed difference between the nominal ($-8e$)  and effective ($\approx -7e$) values of $Q_f$ for LEDWAS (see also \cite {Kaufman:16}.

\addcontentsline{toc}{section}{Conclusion}
\section*{Conclusions} \label{sec:conclusions}

We have reported the initial results of the first biological experiments undertaken to test the predictions of the ICB model of ion channel conduction. In particular, we used patch-clamp experiments to investigate Ca$^{2+}$/Na$^+$ conduction and selectivity, AMFE, and ionic concentration dependences in the bacterial NaChBac channel ($Q_f=-4e$) and in its charge-varied mutants with $Q_f = 0$ and $Q_f=-8e$. The results are compared with earlier Brownian dynamics simulations of the permeation process and with theoretical predictions of the ICB model, which we have extended to encompass bulk concentration affects.

We find that the ICB model provides a good account of both the experimental (AMFE and valence selectivity) and the simulated (discrete multi-ion conduction and occupancy band) phenomena observed in Ca$^{2+}$ channels, including concentration-related shifts of conduction/occupancy bands. In particular we have shown that growth of $Q_f$ from $-4e$ to $-8e$ leads to strong divalent blockade of the sodium current by micromolar concentrations of Ca$^{2+}$ ions, similar to the effects seen in calcium channels. The onset of divalent blockade (shape of the current-concentration curve) follows the Langmuir isotherm, consistent with ICB model predictions.

\addcontentsline{toc}{section}{Acknowledgements}
\section*{Acknowledgements}

The authors gratefully acknowledge valuable discussion with Igor Khovanov, Carlo Guardiani and Aneta Stefanovska.  The research was supported by the UK Engineering and Physical Sciences Research Council grant No.\  	EP/M015831/1, ``Ionic Coulomb blockade oscillations and the physical origins of permeation, selectivity, and their mutation transformations in biological ion channels''.

\addcontentsline{toc}{section}{Appendix. Generation, expression and measurements of NaChBac channels}
\section*{Appendix 1. Generation, expression and measurements of NaChBac channels} \label{subsec:generation}

NaChBac (GenBank accession number BAB05220 \cite{Ren:01}) cDNA was synthesised by EPOCH Life Science (www.epochlifescience.com) and subcloned into the mammalian cell expression vector pTracer-CMV2 (Invitrogen). Amino acid mutations in the pore region of NaChBac were introduced using the Q5$^{\textregistered}$ SDM Kit (New England BioLabs Inc.) in accordance with the manufacturer’s instructions. All mutations were confirmed by DNA sequencing prior to transfection of Chinese Hamster Ovary (CHO) cells with TransIT-2020 (Mirus Bio). Transfected cells (expressing GFP) were identified with an inverted fluorescence microscope (Nikon TE2000-s) and their electrophysiological properties were determined 24--48 hours after transfection.

Whole-cell currents were recorded using an Axopatch 200A (Molecular Devices, Inc., USA) amplifier. Patch clamp signals were digitized using Digidata1322 (Molecular Devices, Inc., USA) and filtered at 2 kHz.  Patch-clamp electrodes were pulled from borosilicate glass (Kimax, Kimble Company, USA) and exhibited resistances of 2–3 MOhm. The shanks of the pipettes tip were coated with beeswax in order to reduce pipette capacitance. The pipette (intracellular) solution contained (in mM): 120 Cs-methanesulfonate, 20 Na-gluconate, 5 CsCl, 10 EGTA, and 20 HEPES, pH 7.4 (adjusted by CsOH). Giga-Ohm seals were obtained in the bath (external) solution containing (in mM): 140 Na-methanesulfonate, 5 CsCl, 10 HEPES and 10 glucose, pH 7.4 (adjusted by CsOH), in which Na-methanesulfonate then was subsequently replaced with Ca-methanesulfonate in order to vary Na$^+$ and Ca$^{2+}$ solution content (see main text). We used methanesulfonate salts in solutions to diminish the influence of endogenous chloride channels. Solutions were filtered with a 0.22 mm filter before use. Osmolarity of all solutions was 280 mOsm (adjusted using sorbitol).

Current–voltage data were typically collected by recording responses to a consecutive series of step pulses from a holding potential of $-100$ mV at intervals of 15 mV beginning at +95 mV. The bath solution was grounded using a 3M KCl agar bridge. 
All experiments were conducted at room temperature.



\addcontentsline{tic}{section}{References}

\bibliography{}        

%
%

\end{document}